\documentclass[aps,prd,twocolumn,superscriptaddress,amsmath,amssymb,eqsecnum,amsthm,nofootinbib, preprintnumbers]{revtex4-2}
\usepackage{xcolor}
\usepackage{url}
\usepackage[colorlinks=true,
    linkcolor=blue,
    citecolor=blue,
    urlcolor=blue]{hyperref}
\usepackage{subfigure}
 \usepackage{amsmath}
\usepackage{graphicx}
\usepackage{epstopdf}
\usepackage{float}
\usepackage{hyperref}
\usepackage{color}
\usepackage[T1]{fontenc}
\usepackage[utf8]{inputenc}
\usepackage[toc,page]{appendix}
\usepackage[usenames,dvipsnames]{xcolor}
\usepackage[normalem]{ulem}

\usepackage{tikz}
\usetikzlibrary{shapes,positioning,shadows,graphs.standard,automata,arrows}
\usepackage{pgfplots}
\pgfplotsset{compat=newest, width=2.669cm, height=2.669cm, scale only axis=true,enlargelimits=false}
\pgfplotsset{tick label style={font=\tiny}}
\pgfplotsset{every major tick/.append style={major tick length=3pt}}
\pgfplotsset{every minor tick/.append style={minor tick length=1.5pt}}
\usepgfplotslibrary{groupplots}
\usepackage{subcaption}
\usepackage{caption}

\providecommand{\renewoperator}[3]{%
\renewcommand*{#1}{\mathop{#2}#3}}

\makeatletter
\providecommand*{\diff}%
{\@ifnextchar^{\DIfF}{\DIfF^{}}}
\def\DIfF^#1{%
\mathop{\mathrm{\mathstrut d}}%
\nolimits^{#1}\gobblespace}
\def\gobblespace{%
\futurelet\diffarg\opspace}
\def\opspace{%
\let\DiffSpace\!%
\ifx\diffarg(%
\let\DiffSpace\relax
\else
\ifx\diffarg[%
\let\DiffSpace\relax
\else
\ifx\diffarg\{%
\let\DiffSpace\relax
\fi\fi\fi\DiffSpace}

\renewoperator{\Re}{\mathrm{Re}}{\nolimits}
\renewoperator{\Im}{\mathrm{Im}}{\nolimits}

\usepackage{bm}

\newcommand{\be}{\begin{equation}}
\newcommand{\ee}{\end{equation}}
\newcommand{\ba}{\begin{eqnarray}}
\newcommand{\ea}{\end{eqnarray}}

\newcommand{\beq}{\begin{equation}}
\newcommand{\eeq}{\end{equation}}
\newcommand{\beqa}{\begin{eqnarray}}
\newcommand{\eeqa}{\end{eqnarray}}

\usepackage[multiple]{footmisc}

\DeclareMathOperator\arctanh{arctanh}

\usepackage{xcolor}
\usepackage{scalerel}

\usepackage{tikz}
\usetikzlibrary{svg.path}
\definecolor{darkmagenta}{HTML}{8B008B}
\definecolor{orcidlogocol}{HTML}{A6CE39}
\tikzset{
  orcidlogo/.pic={
    \fill[orcidlogocol] svg{M256,128c0,70.7-57.3,128-128,128C57.3,256,0,198.7,0,128C0,57.3,57.3,0,128,0C198.7,0,256,57.3,256,128z};
    \fill[white] svg{M86.3,186.2H70.9V79.1h15.4v48.4V186.2z}
                 svg{M108.9,79.1h41.6c39.6,0,57,28.3,57,53.6c0,27.5-21.5,53.6-56.8,53.6h-41.8V79.1z M124.3,172.4h24.5c34.9,0,42.9-26.5,42.9-39.7c0-21.5-13.7-39.7-43.7-39.7h-23.7V172.4z}
                 svg{M88.7,56.8c0,5.5-4.5,10.1-10.1,10.1c-5.6,0-10.1-4.6-10.1-10.1c0-5.6,4.5-10.1,10.1-10.1C84.2,46.7,88.7,51.3,88.7,56.8z};
  }
}

\newcommand\orcidicon[1]{\href{https://orcid.org/#1}{\mbox{\scalerel*{
\begin{tikzpicture}[yscale=-1, scale=0.05, transform shape]
\pic{orcidlogo};
\end{tikzpicture}
}{|}}}}

\begin{document}


\title{Melvin--Bonnor and Bertotti--Robinson spacetimes with Baryonic charge}

\author{Jos\'e Barrientos\,$^{\orcidicon{0000-0003-3445-8151}}$}
\email{jbarrientos@academicos.uta.cl}
\affiliation{Sede Esmeralda, Universidad de Tarapac{\'a}, Avenida Luis Emilio Recabarren 2477, Iquique, Chile}
\affiliation{Institute of Mathematics of the Czech Academy of Sciences, {\v Z}itn{\'a} 25, 115 67 Praha 1, Czech Republic}
\affiliation{Vicerrector\'ia de Investigaci\'on y Postgrado, Universidad de La Serena, La Serena 1700000, Chile}
\author{Fabrizio Canfora\,$^{\orcidicon{0000-0002-4661-9875}}$}
\email{fabrizio.canfora@uss.cl}
    \affiliation{Centro de Estudios Cient\'ificos (CECs), Avenida Arturo Prat 514, Valdivia, Chile}
    \affiliation{Facultad de Ingenier\'ia, Universidad San Sebasti\'an, sede Valdivia, General Lagos 1163, Valdivia 5110693, Chile}
\author{Adolfo Cisterna\,$^{\orcidicon{0000-0002-6949-7882}}$}
\email{adolfo.cisterna.r@mail.pucv.cl}
\affiliation{Sede Esmeralda, Universidad de Tarapac{\'a}, Avenida Luis Emilio Recabarren 2477, Iquique, Chile}
\author{Keanu M{\"u}ller\,$^{\orcidicon{0009-0007-1048-1847}}$}
\email{keanumuller2016@udec.cl}
\affiliation{Departamento de F\'isica, Universidad de Concepci\'on,
Casilla, 160-C, Concepci\'on, Chile}
\author{Anibal Neira\,$^{\orcidicon{0009-0004-6593-6093}}$}
\email{aneira2017@udec.cl}
\affiliation{Departamento de F\'isica, Universidad de Concepci\'on,
Casilla, 160-C, Concepci\'on, Chile}



\begin{abstract}
In this work, we exploit a dictionary that maps solutions of the Einstein--Scalar--Maxwell theory to those of gauged Skyrme--Maxwell--Einstein models in $(3+1)$ dimensions, allowing gravitating scalar-field configurations in external electromagnetic backgrounds to be interpreted in terms of baryonic quantities. Using the definition of Baryonic charge as the integral of a topological density, we derive novel mass formulas that relate the spacetime mass parameter to the Baryonic charge, the external magnetic field, and other solution parameters. As a result, the mass and Baryonic charge are not independent.
These closed analytic expressions encode physical information that would otherwise be difficult to extract. In particular, for the class of solutions considered here, the relation between mass and Baryonic charge is linear at large mass, while significant nonlinear deviations appear at intermediate values. This framework provides a useful tool for both extracting new physical insights from known solutions and constructing new baryonic configurations using the solution-generating techniques available for Einstein--Scalar--Maxwell theory.

\end{abstract}

\maketitle
\section{Introduction}

There is little doubt that some of the most important—and at the same time most challenging—open problems in $(3+1)$-dimensional General Relativity (GR), cosmology, and astrophysics are related to the dynamics of baryonic matter in regions of strong gravity, particularly in the presence of additional ingredients such as stationarity of the compact source or intense electromagnetic fields \cite{Broderick:2001qw,Sathyaprakash:2009xs,Giataganas:2018uuw,Watanabe:2020vas,Berryman:2022zic}.
In such regimes, numerical simulations are not only computationally demanding, but the available analytical tools are also insufficient to construct exact black hole solutions carrying Baryonic charge. Yet the pursuit of analytic solutions is well motivated: as illustrated historically by the discovery of the Schwarzschild and Kerr geometries, exact configurations often reveal novel physical phenomena that would be difficult to uncover by other means.

The technical difficulty in coupling GR to magnetized baryonic matter stems from the simultaneous presence of strong nonlinearities in both GR and in the low-energy limit of QCD. The latter is itself notoriously challenging, since neither lattice QCD nor perturbative methods are reliable in this regime.
Nevertheless, the low energy limit of QCD admits an effective description in terms of $(3+1)$-dimensional gauged Skyrme--Maxwell theory (GSMT) with $SU(2)$ isospin symmetry \cite{Skyrme:1961vr,Skyrme:1961vq,Skyrme:1962vh,Witten:1983tx,Balachandran:1982ty,Adkins:1983ya,Balachandran:1991zj}. Within this framework, GSMT occupies a central role in chiral perturbation theory; see \cite{Scherer:2012xha,Donoghue:1992dd,Machleidt:2011zz,Gasser:1983yg,Leutwyler:1993iq,Ecker:1994gg}, where the associated topological charge density is naturally identified with the Baryonic charge density.
The development of effective analytical tools in this setting is therefore highly desirable. Such tools would not only complement and inform numerical simulations but would also provide means to reveal genuinely new phenomena associated with highly magnetized baryonic matter in regimes of strong gravitational fields.

Recently, a mapping relating GSMT and the Einstein--Scalar--Maxwell theory—and, consequently, their respective spectra of exact solutions—has been established \cite{Canfora:2026wfi}. This correspondence allows one to systematically ``transfer'' solutions of the Einstein--Scalar--Maxwell system, which carry no Baryonic charge, into the GSMT sector, where they acquire a nontrivial Baryonic charge profile.
While constructing solutions in the Einstein--Scalar--Maxwell theory is nontrivial, the framework admits a wide range of well-developed solution-generating techniques. The key advantage of the mapping is that the full effectiveness of these techniques can be directly inherited by the GSMT sector, thereby greatly enlarging the class of analytically accessible configurations with Baryonic charge.

Here, after a concise review of the mapping, we apply it to the case in which the seed spacetimes to be translated into the GSMT sector are Melvin--Bonnor \cite{Bonnor:1954tis,Ernst:1976mzr,Barrientos:2024pkt} and Bertotti--Robinson geometries \cite{Bertotti:1959pf,Robinson1,Podolsky:2025tle,Alekseev:1996fq}, namely configurations describing fully backreacting electromagnetic fields, with or without an embedded compact source. Since the presence of a scalar field is a crucial ingredient of the dictionary, we first dress these Melvin--Bonnor and Bertotti--Robinson backgrounds with an appropriate scalar field profile, thereby rendering them suitable seeds. This is achieved through the Eris--Gürses theorem \cite{Eris:1976xj}, a solution-generating technique valid in the electrovacuum.
With these ingredients in place, the baryonic sector can be analyzed in a systematic manner, and the corresponding Baryonic charge can be explicitly computed. Remarkably, a natural quantization condition emerges, relating the Baryonic charge to the parameters characterizing the original seed spacetime.

This work is organized as follows. In \autoref{sec2}, we review the main ingredients of the mapping introduced in \cite{Canfora:2026wfi}. \autoref{sec3} is devoted to presenting the necessary elements for constructing the seed configurations and to combining them in order to explicitly build the corresponding solutions on the GSMT side, where the associated Baryonic charges are computed. We analyze their behavior and derive the resulting quantization conditions. Finally, in \autoref{sec4}, we present our concluding remarks and outline several directions for future investigation.

\section{The Mapping}\label{sec2}

Our starting point is the action of GSMT, which is given by (see
\cite{Canfora:2026wfi})
\begin{widetext}
\begin{align} S=\int d^{4}x \sqrt{-g} \, \left( \frac{1}{4} R -\frac{1}{4}F_{\mu\nu} F^{\mu\nu}
+\frac{K}{4}\mathrm{Tr}\left[ \Sigma^{\mu}\Sigma_{\mu}+\frac{\lambda} {8}B_{\mu\nu}B^{\mu\nu}\right] \right).\label{ESMaction}
\end{align}
\end{widetext}
Here, $R$ denotes the Ricci scalar and $K=(f_{\pi})^{2}/4$ defines the Skyrme coupling constant, which is experimentally fixed to $f_{\pi}\approx130\ \text{MeV}$. Note that we adopt the convention $c=4\pi G=\epsilon_{0}=1$. On the other hand, the pion mass $m_{\pi}$ is neglected, as we will focus on configurations with large Baryonic charge; thus, $m_{\pi}$ is much smaller than the characteristic energy scales involved.

As usual, the Einstein field equations are 
\begin{align}
R_{\mu\nu} - \frac{1}{2}g_{\mu\nu}R = 2 T_{\mu\nu}, \label{EinsteinEqu1}%
\end{align}
where the matter source is given by 
\begin{widetext}
\begin{equation}
T_{\mu\nu}=  -\frac{K}{2}\operatorname{Tr}\bigg[ \Sigma_{\mu}\Sigma_{\nu
}-\frac{1}{2}g_{\mu\nu}\Sigma_{\sigma}\Sigma^{\sigma}
 +\frac{\lambda}{4}\left(  g^{\alpha\beta}B_{\mu\alpha}B_{\nu\beta}-\frac
{1}{4}g_{\mu\nu}B_{\alpha\beta}B^{\alpha\beta}\right)  \bigg]
 + F_{\mu\alpha}F_{\nu}^{\;\alpha}-\frac{1}{4}F_{\alpha\beta}F^{\alpha\beta} g_{\mu\nu}. \label{tmunu(1)}
\end{equation}
\end{widetext}

Here
\begin{align}
\Sigma_{\mu}  &  =U^{-1}D_{\mu}U=\Sigma_{\mu}^{j}t_{j},\quad F_{\mu\nu}%
=\partial_{\mu}A_{\nu}-\partial_{\nu}A_{\mu},\nonumber\\
B_{\mu\nu}  &  =\left[  \Sigma_{\mu},\Sigma_{\nu}\right],\quad t_{j}%
=i\sigma_{j},
\end{align}
where $\sigma_{i}$ are the Pauli matrices and the $SU(2)$ field $U$ is parametrized by
\begin{align}
U  &  =1_{2\times2}\cos\alpha+\left(  \sin\alpha\right)  n^{j}t_{j}%
,\nonumber\\
\overrightarrow{n}  &  =\left(  \sin\Theta\cos\Phi,\sin\Theta\sin\Phi
,\cos\Theta\right), \label{eq:defSigma}
\end{align}
with $1_{2\times2}$\ denoting the identity $2\times2$ matrix while $\alpha\left(
x^{\mu}\right)  $, $\Theta\left(  x^{\mu}\right)  $ and $\Phi\left(  x^{\mu
}\right)  $ are the three scalar degrees of freedom of the $SU(2)$-valued
Chiral field. On the other hand, the gauge-covariant derivative is 
\begin{equation}
D_{\mu}U=\partial_{\mu}U+A_{\mu}\left[  t_{3},U\right]. \label{sky2.75}%
\end{equation}
Now, the field equations for the gauge and Skyrme fields are 
\begin{align}
D_{\mu}\left(  \Sigma^{\mu}+\frac{\lambda}{4}[\Sigma_{\nu},B^{\mu\nu}]\right)
=0, \quad\nabla_{\mu}F^{\mu\nu}=J^{\nu}, \label{eq:NLSM}%
\end{align}
being $\nabla_{\mu}$ the Levi-Civita covariant derivative, and the current
$J^{\mu}$ defined by 
\begin{equation}
J^{\mu}=\frac{K}{2}\text{Tr}\left[  \widehat{O}\left(  \Sigma^{\mu}%
+\frac{\lambda}{4}[\Sigma_{\nu},B^{\mu\nu}]\right)  \right],
\label{maxcurrent}%
\end{equation}
where $\widehat{O}=U^{-1}t_{3}U-t_{3}$.

The Baryonic charge \cite{Skyrme:1961vr,Skyrme:1961vq,Skyrme:1962vh,Witten:1983tx,Balachandran:1982ty,Adkins:1983ya,Balachandran:1991zj}
\begin{equation}
Q_B=\frac{1}{24\pi^{2}}\int_{\mathcal{H}}\rho_{B}, \quad\rho_{B}=\rho_{B_1}+\rho_{B_2},
\label{new4.1}%
\end{equation}
with $\mathcal{H}$ being a spacelike hypersurface of dimension three, it is composed of the two distinct topological density terms
\begin{align}
\rho_{B_1}  &  =\epsilon^{ijk}\text{Tr}\left[  \left(  U^{-1}\partial
_{i}U\right)  \left(  U^{-1}\partial_{j}U\right)  \left(  U^{-1}\partial
_{k}U\right)  \right],\label{rhoSk}\\
\rho_{B_2}  &  =-3\epsilon^{ijk}\text{Tr}\left[  \partial_{i}\left[
A_{j}t_{3}\left(  U^{-1}\partial_{k}U+\left(  \partial_{k}U\right)
U^{-1}\right)  \right]  \right], \label{rhoMax}%
\end{align}
where $A_{j}$ denotes the spatial components of the gauge potential. The term $\rho_{B_{1}}$ corresponds to the standard Skyrme contribution, while the second term, $\rho_{B_{2}}$, is the so-called Callan--Witten contribution \cite{Callan:1983nx}.

Keeping these standard technical details in mind, we can formulate a first observation crucial to the mapping introduced in \cite{Canfora:2026wfi} and leading to the desired simplification of the GSMT equations. It is possible to construct a topologically nontrivial ansatz for the Skyrme field for which the Baryonic charge is entirely supported by the Callan--Witten term alone, namely by $\rho_{B_2}\neq 0$, while the standard Skyrme contribution vanishes. In this case, the hadronic and electromagnetic degrees of freedom are drastically simplified, which, in turn, greatly facilitates the construction of gravitating configurations carrying nontrivial Baryonic charge. In mathematical terms, the following ansatz is considered for the $SU(2)$-valued Skyrme and gauge fields $U$ and $A$ 
\begin{align}
\Psi & =\Psi(x^{\mu}),\quad  \Theta=\pi ,\quad  \Phi=0\Rightarrow\ U=\exp\left(
\Psi t_{3}\right),\nonumber \\
 A_{\nu}&=A_{\nu}(x^{\mu})\label{LWP0}.
\end{align}
Following the notation of \cite{Canfora:2026wfi}, the Skyrme profile usually denoted by $\alpha$ is here denoted as $\Psi$. 

It is precisely for this ansatz that the Skyrme and Maxwell equations \eqref{eq:NLSM} simplify to the Klein--Gordon equation and the sourceless Maxwell equations, respectively\footnote{A restriction arises from the low-energy approximation inherent in GSMT. This effective description of QCD ceases to be valid once the energy density associated with the hadronic degrees of freedom becomes too large. In the present context, this translates into imposing  \begin{equation}
\left\vert \nabla_{\nu}\Psi\right\vert \lesssim1 \text{GeV}. \label{LQCD}%
\end{equation}}$^{,\,}$\footnote{To understand the underlying details of this simplification, we refer to \cite{Canfora:2026wfi}.}
\begin{align}
\Box\Psi=0,\quad \nabla_{\mu}F^{\mu\nu}  =0.
\label{LWP2}%
\end{align}

An important aspect of this construction is the role of the ansatz for the chiral field $U$. This ansatz restricts the dynamics to the neutral pion sector, effectively decoupling the Maxwell equations while preserving a nonvanishing topological charge density. Similar configurations have been considered in the literature (see, for instance, \cite{Brauner:2016pko, Moss:2000hf}, and references therein).
The distinctive feature of the present approach lies not in the ansatz itself, but in its combination with the solution-generating techniques available in the Einstein--Scalar--Maxwell framework, such as the Ernst equation method \cite{Ernst:1967wx,Ernst:1967by} and the Lie point symmetries it discloses in a very intuitive manner. This framework enables the generation of solutions with nonvanishing Baryonic charge starting from seed configurations with vanishing Baryonic charge. Within this perspective, gravitational backgrounds—independently of their novelty from a purely general relativistic standpoint—acquire a new physical interpretation as solutions of the Einstein–Gauged–Skyrme–Maxwell theory endowed with Baryonic charge.
A key outcome of this approach is the ability to relate the Baryonic charge to the solution's parameters. In particular, one can derive mass formulas for magnetized black holes in terms of the Baryonic charge and the external magnetic field. These relations follow directly from the expression for the topological charge density and provide access to physical information that is otherwise difficult to extract. To the best of the author’s knowledge, such mass formulas have not been previously reported in the literature.
More generally, this framework allows one both to construct new configurations via the Ernst method and to endow known geometries with Baryonic charge in a manner consistent with the gauged Skyrme model. This relies on the fact that, under the chosen ansatz, the hadronic degrees of freedom are effectively mapped to a free-field description with nonvanishing Baryonic charge density. As a result, one can obtain novel insights into how quantities such as the black hole mass or angular momentum depend on the Baryonic charge.

\section{Baryonic Melvin--Bonnor and Bertotti--Robinson solutions}\label{sec3}

As stated above, in order to construct solutions of the fully backreacting GSMT, it is sufficient to first obtain solutions of the Einstein--Scalar--Maxwell theory and then, via the mapping introduced in \cite{Canfora:2026wfi}, identify the scalar field that plays the role of the Skyrme degree of freedom. The nontrivial character of the Baryonic density charge crucially relies on the presence of a magnetic field suitably aligned with the gradient of the scalar profile.

Exact solutions endowed with magnetic fields fall into two broad classes: those with localized magnetic charge, such as magnetic monopoles, and those with delocalized magnetic fields, hereafter referred to as external magnetic fields. Our interest lies in the latter case. In particular, we will construct solutions featuring two types of external magnetic fields: Melvin--Bonnor and Bertotti--Robinson.

We begin with the better-known Melvin--Bonnor spacetimes. In electrovacuum and in the absence of compact sources, the Melvin--Bonnor solution describes a static, cylindrically symmetric spacetime representing the backreaction of a magnetic field sharing the same symmetries. It can be interpreted as a magnetic flux tube held together by its own gravitational field. Although this solution can be obtained by direct integration of the field equations, it is more commonly understood as arising from the application of a Harrison transformation \cite{Harrison} to Minkowski spacetime. This approach is advantageous, as direct integration fails in more general settings. The Harrison transformation is a Lie point symmetry of the electrovacuum equations that becomes manifest in the Ernst complex potential formulation \cite{Ernst:1967wx,Ernst:1967by}. Its generic effect is the generation of electromagnetic fields, whose nature—localized or delocalized—depends on the choice of Killing vector of the seed spacetime used in the transformation.

In the present context, however, the seed spacetime must also support a nontrivial scalar field. Direct integration of the coupled field equations in this case is highly nontrivial. Fortunately, several techniques exist to consistently dress electrovacuum configurations with scalar fields. The most general among them is the Eris--Gürses theorem \cite{Eris:1976xj}, which has the additional advantage of commuting with charging transformations.

To keep the construction as simple as possible, we will consider Minkowski and Schwarzschild spacetimes as vacuum seeds, dress them with appropriate scalar field profiles, and subsequently magnetize them. These configurations will serve as the starting point for computing the Baryonic charge in the GSMT framework. Here, we only outline the main steps of the construction, referring the reader to \cite{Canfora:2026wfi} for the detailed derivation.

Given any stationary and axisymmetric electrovacuum configuration, the backreaction of a scalar field sharing the same symmetries can be straightforwardly incorporated by means of the Eris--Gürses theorem. To make this procedure explicit, we focus on the static vacuum case—which will be sufficient for our purposes, as we shall consider Minkowski and Schwarzschild spacetimes in what follows—and write the line element in canonical Weyl--Lewis--Papapetrou coordinates $\{t,\rho,z,\varphi\}$, where $-\infty<t<\infty$ and the coordinates $\{\rho,z,\varphi\}$ take their standard cylindrical ranges
\begin{equation}
    ds^2=-e^{2U}dt^2+e^{-2U}\left[e^{2\gamma}(d\rho^2+dz^2)+\rho^2 d\varphi^2\right],
\end{equation}
with $U$ and $\gamma$ functions of $\{\rho,z\}$ only.
For such a line element, the field equations of the Einstein--Scalar system naturally decouple the backreaction of the scalar. In fact, given a vacuum solution $(U,\gamma)$ a solution of the Einstein--Scalar theory is given by $(U,\bar{\gamma}=\gamma+\gamma^{\Psi},\Psi)$ where the scalar backreaction is found via 
\begin{equation}
\gamma_{, \rho}^{\Psi}=\rho\left(\Psi_{, \rho}^2-\Psi_{, z}^2\right), \quad \gamma_{, z}^{\Psi}=2 \rho \Psi_{, \rho} \Psi_{, z}, \quad \nabla_{\mathbb{E}^3}^2 \Psi=0 .
\end{equation}
This implies that any vacuum solution can be immersed in a scalar multipolar background determined by the above quadratures, once a particular solution of the Laplace equation is chosen. The general solution of the Laplace equation contains two distinct types of contributions: asymptotically flat terms, which decay as negative powers of $\mathcal{R}=\sqrt{\rho^{2}+z^{2}}$, and growing terms, which instead behave as positive powers of $\mathcal{R}$ at large distances.

By direct inspection, one finds that within the latter (growing) branch, it is possible to identify configurations that give rise to a nontrivial Baryonic charge. Motivated by this observation, we will restrict our attention to the simplest scalar multipolar background, namely that associated with the first term in the growing sector of the Laplace equation solutions.
Taking the Schwarzschild spacetime as the seed (with Minkowski recovered in the massless limit), the corresponding Einstein--Scalar configuration, after transforming back to spherical coordinates, reads \cite{Cardoso:2024yrb} 
\begin{equation}\label{natario}
ds^2=-fdt^2+H\left[\frac{dr^2}{f}+r^2d\theta^2\right]+r^2\sin^2\theta d\varphi^2,
\end{equation}
with
\begin{equation}\label{natario}
\begin{aligned} 
H(r,\theta)=e^{-{K^2}r^2f\sin^2\theta},\quad\Psi(r,\theta)={K}(r-M)\cos\theta,
\end{aligned}
\end{equation}
{where $K$ is the scalar charge, and  $f(r)=1-\frac{2{M}}{r}$.} With this at hand, the magnetization procedure becomes straightforward. Following \cite{Dowker:1993bt}, where it has been shown at the metric level that, for a given axisymmetric vacuum (or scalar-dressed) solution of the form $ds^2=g_{ij}dx^idx^j+g_{\varphi\varphi}d\varphi^2$, its magnetized counterpart is obtained as
$ds^2=\Lambda^2(g_{ij}dx^idx^j)+g_{\varphi\varphi}d\varphi^2/\Lambda^2$ together with
$A=Bg_{\varphi\varphi}d\varphi/2\Lambda$, where $\Lambda=1+B^2g_{\varphi\varphi}/4$, we are led to the configuration 
\begin{align}\label{schwmelvinmelvin}
ds^2&=\Lambda^2\left(-fdt^2+H\left[\frac{dr^2}{f}+r^2d\theta^2\right]\right)+\frac{r^2\sin^2\theta}{\Lambda^2}d\varphi^2,\nonumber\\
 A&=-\frac{B}{2}\frac{r^2\sin^2\theta}{\Lambda} d\varphi,
\end{align}
where $\Lambda(r,\theta)=1+\frac{B^{2}}{4}r^2\sin^2\theta$. The scalar profile remains unchanged.

Some of the main features of this configuration have already been discussed in \cite{Cardoso:2024yrb}. However, for completeness, we highlight here some of its most relevant aspects and complement the discussion with a few interesting limits. First, one can note that the geometry described by the metric \eqref{schwmelvinmelvin} is of Petrov type I, in close analogy with the Schwarzschild--Melvin solution. The spacetime is not asymptotically flat due to the presence of the electromagnetic field, parametrized by $B$, which persists at any value of $z$, because the scalar field gradient becomes aligned with the $z$-axis at large distances. In addition, the scalar profile grows with the radial coordinate (more precisely, along the original coordinate $z$), which leads to the appearance of a singularity at infinity. As usual, the central singularity is associated with the compact source at the origin.
The singular behavior can be seen explicitly from the expression of the Kretschmann invariant, both at the origin and at infinity
\begin{equation}
\begin{aligned}
R_{\mu\nu\rho\sigma}R^{\mu\nu\rho\sigma}&\underset{r\rightarrow 0}{\sim}\frac{48M^2}{r^6},\\
R_{\mu\nu\rho\sigma}R^{\mu\nu\rho\sigma}&\underset{r\rightarrow \infty}{\sim}e^{2K^2r^2f\sin^2\theta}.
\end{aligned}
\end{equation}
For {$K=0$}, the solution reduces to the Schwarzschild-Melvin geometry, whereas for $M=0$ the spacetime remains Petrov type I, unless $B$ or {$K$} vanish. This background, obtained in the massless limit, will prove useful in what follows, as it allows for a regularization of the Baryonic charge. 

The physical intuition is as follows. The Melvin spacetime with the Hadronic profile introduced above is a background that possesses a very large number of Baryons. However, such particles are distributed homogeneously in such a way that no black hole is formed. On the other hand, if extra Baryons are added to the Baryonic Melvin background (creating a Baryon density peak somewhere), then a black hole may form. This configuration would be a Baryonic black hole in a Melvin background. The important quantity is then the Baryonic charge excess due to the presence of the black hole. In other words, the relevant observable is the difference between the Baryonic charge of the Baryonic Melvin black hole and the Baryonic charge of the Baryonic Melvin background (without black hole). Such a quantity encodes the dependence of the black hole mass parameter on the magnetic field and on the Baryonic charge. One of the main results of the present work is that we can compute this dependence analytically.

\begin{figure}[H]
    \centering
\includegraphics[width=1.0\linewidth]{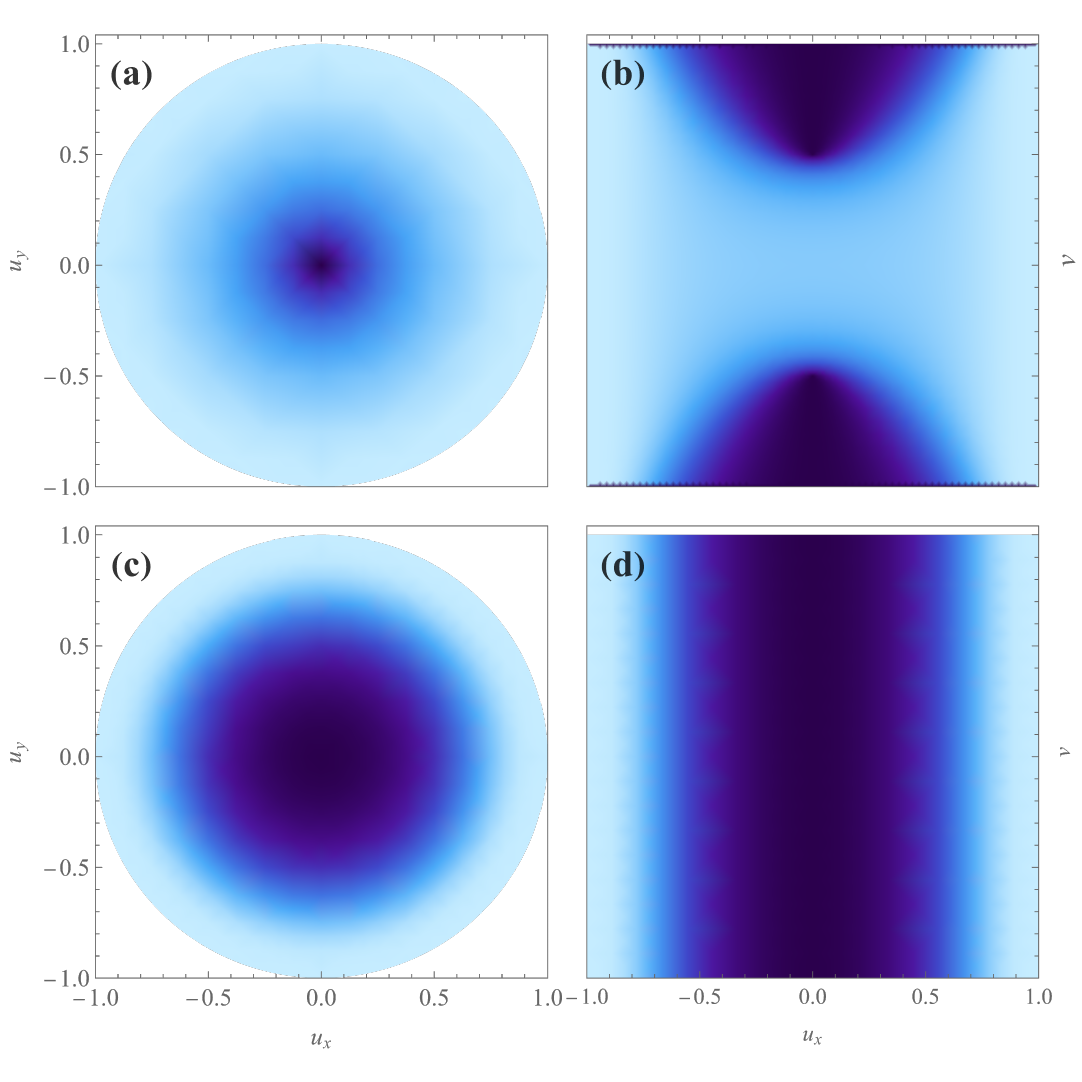}
    \caption{In panel (a), a $v=\mathrm{const.}$ slice of the Schwarzschild--Melvin configuration is displayed. Panel (b) shows a slice at $u_y=0$. Panels (c) and (d) present the same sections for the Melvin background solution $(M=0)$. Regions of higher color intensity correspond to a higher Baryonic charge density.}
    \label{fig1}
\end{figure}

{Since the Baryonic charge density is a topological density, it is independent of the local form of the metric. Consequently, the Baryonic charge itself is a topological invariant. This can be seen explicitly from the fact that the factor $\sqrt{-g}$ appearing in the volume element cancels the corresponding factor in the denominator of the Levi-Civita tensor. Therefore, when evaluating the topological density, one may equivalently work with a flat metric. As it has already been stated in our setup, the Baryonic charge is entirely determined by the Callan--Witten term $\rho_{B_2}$, which in the present context reduces to 
\begin{equation}
\rho_{B_2}=\epsilon_{ijk}\partial_iA_j\partial_k\Psi.
\end{equation}
To compute it for the configuration \eqref{schwmelvinmelvin}, we consider Euclidean three-dimensional flat spacetime adapted to the problem, namely, expressed in spherical coordinates. A direct computation then yields 
\begin{equation}\label{rhob21}
\rho_{B_2}=\frac{K B(r-M\sin^2\theta)}{r\Lambda^2}.
\end{equation}}
The spatial distribution of the topological charge density $\rho_{B_2}$ can be seen in \autoref{fig1}.

In this case, one observes that along the $\rho$ direction, the density is concentrated near the black hole horizon, with darker blue regions indicating higher density. However, along the $z$ direction, the density increases as one approaches infinity, $|v|=1$, as reflected by the darker blue shading. This behavior indicates a probable divergence of the topological charge itself.

To plot the Baryonic charge density, we have proceeded as follows: we first change to coordinates  
\begin{widetext}
\begin{equation}
   r=M+\sqrt{\frac{\rho^2+z^2+M^2+\sqrt{(\rho^2+z^2+M^2)^2-4M^2 z^2}}{2}}, \qquad \cos\theta=\frac{\sqrt{2}z}{\sqrt{\rho^2+z^2+M^2+\sqrt{(\rho^2+z^2+M^2)^2-4M^2 z^2}}}.
\end{equation}
\end{widetext}
In this chart, the horizon surface is defined as a closed line segment of length $2M$ on the $z$-axis (with its center at $z=0$). One can formally approach it for $|z|\leq M$ by taking $\rho\to 0$. Then, one can employ the convenient coordinates 
\begin{equation*}
(u_{\mathrm{x}},u_{\mathrm{y}})=\frac{2}{\pi}\arctan \rho\,(\sin\varphi,\cos\varphi),\quad
    v=\frac{2}{\pi}\arctan z,\label{eq:uvcoords}
\end{equation*}
in which $|z|$ infinity sits at $|v|=1$ and $\rho$ infinity at $u_{\mathrm{x}}^2+u_{\mathrm{y}}^2=1$. The horizon is again understood as a line segment, now of length $(4/\pi)\arctan M$, on the $v$ axis (with center at $v=0$). 

The topological Baryonic charge ${Q}_B$ of the system is obtained by integrating $\rho_{B_2}$ over the volume;\footnote{{As we have emphasized, the metric can be taken as flat when computing the Baryonic charge density due to its topological nature. Thus, the volume element is $dV=r^2\sin\theta dr d\theta d\varphi$.}} 
it yields 
{
\begin{widetext}
\begin{align}\label{eq310}
Q_{B}= \int_{\mathcal{H}} \rho_{B_{2}}=\frac{8\pi{K}}{B^2}\left[B (r-2M)-\frac{4}{\sqrt{4+B^2r^2}}\arctanh\left(\frac{Br}{\sqrt{4+B^2r^2}}\right)+\frac{1}{\sqrt{1+B^2M^2}}\arctanh\left(\frac{BM}{\sqrt{1+B^2M^2}}\right)\right].
   \end{align}
\end{widetext}}
Notice that this topological charge diverges, confirming our intuition, when one considers the entire range $2M<r<\infty$. This divergence arises because both profiles—the scalar field and the magnetic field component $A_\varphi$—are present throughout the whole spacetime and remain constant along the $z$ direction. Note, however, that, as discussed above, this is not surprising at all, since the Baryonic Melvin background contains a very large number of Baryons. What actually matters is the difference between the Baryonic charge of the Baryonic Melvin black hole and the Baryonic Melvin configuration without a black hole. This is exactly the same situation as in the analysis of black hole mass in asymptotically AdS spacetime when the mass is measured with respect to the AdS background.

\begin{figure}[H]
\hspace*{-0.6cm} 
\includegraphics[width=1.1\linewidth]{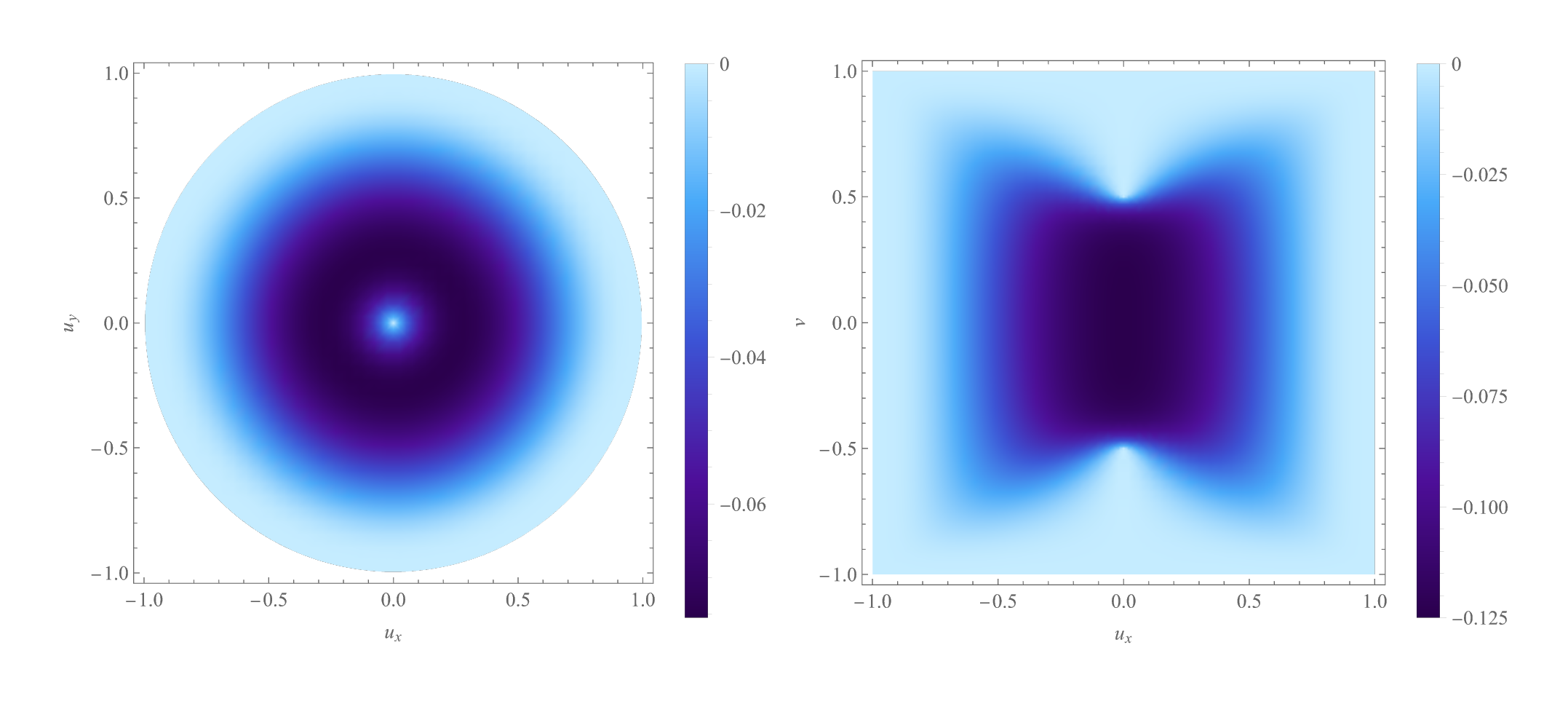}
    \caption{Baryonic charge density obtained after subtracting the contribution of the Melvin background. The darkest regions indicate where the Baryonic charge is strongly concentrated.}
    \label{fig2}
\end{figure}

\begin{figure}[H]
    \centering    \includegraphics[width=1\linewidth]{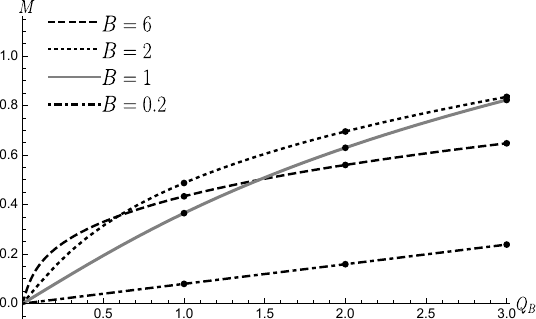}
    \caption{The mass $M$ as a function of the Baryonic charge $Q_B$.}
    \label{fig3}
\end{figure}

{To obtain a finite Baryonic charge, it is therefore necessary to compute the Baryonic charge associated with the background solution $(M=0)$
\begin{equation}
Q_{B}\underset{M=0}{=}\frac{8\pi{K}}{B^2}\left[ B r-\frac{4}{\sqrt{4+B^2r^2}}\arctanh\left(\frac{Br}{\sqrt{4+B^2r^2}}\right)\right],
\end{equation}
and subtract it from \eqref{eq310}, the Baryonic charge corresponding to the configuration in which the compact source is present. The resulting finite Baryonic charge is then given by 
\begin{equation}
Q_{B}=\frac{8\pi{K}}{B}\left[\frac{1}{B\sqrt{1+B^2M^2}}\arctanh\left(\frac{BM}{\sqrt{1+B^2M^2}}\right)-2M\right]. 
\end{equation}}
The corresponding Baryonic density plot is shown in \autoref{fig2}. 
It is important to emphasize the key implication of the formula above: it provides a relation that determines the black hole mass parameter in terms of its (discrete) Baryonic charge and the magnetic field. In principle, one may invert it to obtain the mass as a function of the topological charge and the magnetic field. Although this inversion cannot be carried out analytically in general, since the resulting equation is transcendental in the mass parameter, it is straightforward to plot the mass as a function of the Baryonic charge for different values of the magnetic field (see \autoref{fig3}).

The above transcendental equation can be solved analytically in certain limits of interest. When the mass is small, we get
\begin{equation}
\begin{aligned}
M&=\frac{1}{4}
\frac{\left[\left(-6 Q_B B^{2} + 2 \sqrt{9 B^{4} Q_B^{2} + 128 \pi^{2}{K^2}}\right)\right]^{1/3}}
{\pi^{1/3}{K^{1/3}} B }\\
&\quad-
\frac{2 \pi^{1/3} {K^{1/3}}}
{B\left[\left(-6 Q_B B^{2} + 2 \sqrt{9 B^{4} Q_B^{2} + 128 \pi^{2}{K^{2}}}\right)\right]^{1/3}}.
\end{aligned}
\end{equation}
On the other hand, when the mass is very large, the relationship between the mass and the Baryonic charge becomes linear
\begin{equation}
M=\frac{B}{16\pi{K}}Q_B.
\end{equation}
From this analysis, several relevant pieces of information can be extracted. The most immediate observation is that the mass is expected to be proportional to the Baryonic charge, which represents the number of Baryons. This behavior is indeed recovered in the large-mass regime.

\begin{figure}[h]
    \centering    
    \includegraphics[width=1\linewidth]{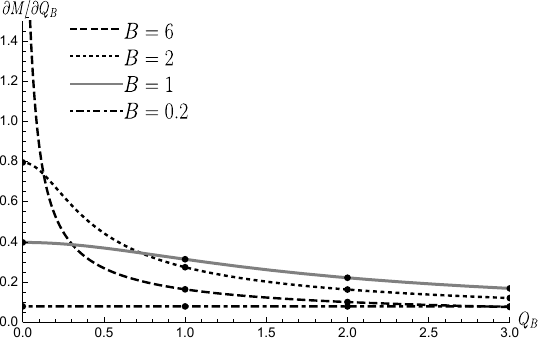}
    \caption{The behavior of the change of $M$ with respect to $Q_B$. For small $B$, $\partial M/\partial Q_B$ becomes constant (dashdotted curve).}
    \label{fig4}
\end{figure}

However, for moderate and small masses, sizeable deviations from this expectation arise. This is precisely the most interesting region, where the interplay between baryonic interactions, the magnetic field, and gravity becomes evident. The plots in \autoref{fig3} and \autoref{fig4} show that the largest departures from linearity occur for strong magnetic fields and small Baryonic charge. In this regime, the additional Baryonic charge required to produce a small increase in mass is significantly lower than in the linear regime. Equivalently, the mass grows more rapidly with Baryonic charge than it does at large mass.

Further information can also be extracted, such as generalized susceptibilities, which are related to derivatives of the mass parameter with respect to the magnetic field. A detailed analysis of these quantities will be presented in future work.

A second exact electromagnetic background of interest is the Bertotti--Robinson spacetime \cite{T.Levi-Civita1,Bertotti:1959pf,Robinson1}. This electrovacuum solution, given by the direct product of two constant-curvature spacetimes, an anti-de Sitter two-dimensional spacetime ($\rm{AdS}_2$) and a two-sphere ($\mathcal{S}^2$), plays an important role not only in the theory of exact solutions of Einstein--Maxwell equations but also in applications to string theory and holography. Embedding a Schwarzschild black hole in the Bertotti--Robinson background again requires the Ernst framework \cite{Alekseev:1996fq}, though in this case, the construction relies not on hidden symmetries but rather on direct integration of the Ernst equations via the so-called monodromy data transform \cite{Alekseev1,Alekseev2,Alekseev3}. 

In direct analogy with the Melvin--Bonnor case analyzed above, our starting point is a configuration describing a Schwarzschild black hole embedded in a Bertotti--Robinson magnetic background, which has already been dressed with a scalar field through the application of the Eris--Gürses theorem
\begin{equation}
\begin{aligned}
&ds^2=\frac{1}{\Omega^2}\left(-Q dt^2+H\left[\frac{dr^2}{Q}+\frac{r^2}{P}d\theta^2\right]+r^2P\sin^2\theta d\varphi^2\right),\\
&\Psi= \frac{{K}}{2}\ln\left(\frac{r^2PQ\sin^2\theta}{\Omega^4}\right),\quad A=\frac{1}{\tilde{b}}\left(\Omega-r\partial_r\Omega-1\right)d\varphi,
\end{aligned}
\end{equation}
where the backreaction of the scalar field is encoded in the function
\begin{equation}
    H(r,\theta)=\left(\frac{r^2PQ\sin^2\theta}{\Omega^4}\right)^{{K^2}}, 
\end{equation}
being 
\begin{equation}
\begin{aligned}
Q(r)&=\frac{\left(r(1-\tilde{b}^2M^2)-2M\right)}{r}\left(1+\tilde{b}^2r^2\right),\\ P(\theta)&=1+\tilde{b}^2M^2\cos^2\theta,\\\Omega(r,\theta)&=\sqrt{1+r\tilde{b}^2\left(r-\left(r(1-\tilde{b}^2M^2)-2M\right)\cos^2\theta\right)}.
\end{aligned}
\end{equation}
Here, $\tilde{b}$ denotes the magnetic field of the Bertotti--Robinson background, while {$K$} is again the scalar charge.

\begin{figure}[h]
    \centering
{\includegraphics[width=0.48\textwidth]{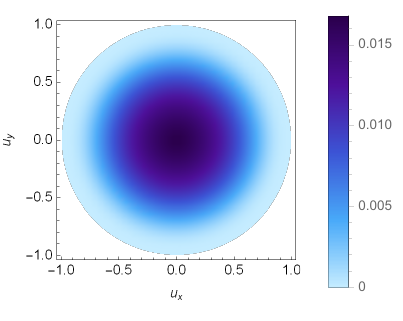}} 
\caption{A $z=\rm{const.}$ slice of $\rho_B$ is displayed, where the charge density is highly concentrated near the event horizon along the $\rho$-direction and decays monotonically as $\rho \to \infty$.}
    \label{fig5}
\end{figure}

\begin{figure}[h]
    \centering
{\includegraphics[width=0.48\textwidth]{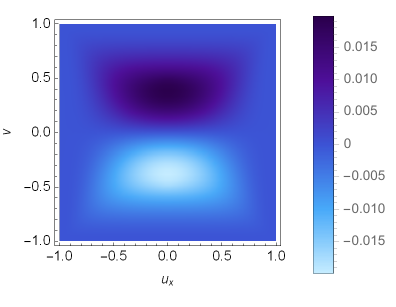}} 
\caption{A $\rho=\rm{const.}$ slice shows that the charge density oscillates between positive and negative values, thereby revealing the polarized nature of the configuration.}
    \label{fig6}
    \end{figure}

The Baryonic charge density associated with the Callan--Witten term is given, in this case, by
\begin{widetext}
\begin{equation}
\rho_{B_2}=\frac{\tilde{b}{K} \cos\theta}{r\Omega^3}\left[\frac{\tilde{b}^2M(r-M)(1+\tilde{b}^2Mr)\cos^2\theta}{(1+\tilde{b}^2r^2)}-\frac{\left(\tilde{b}^2[\tilde{b}^2M^3r^2(\tilde{b}^2Mr+3)-M^2(M-6r)
-r^2(3M-r)]+ M
\right)}{rQ}\right].
\end{equation}
\end{widetext}
The Baryonic charge density shown in \autoref{fig5} and \autoref{fig6} resembles the electric charge distribution that arises when a neutral object develops a ``charge separation'' once an external force acting on the charges is switched on. In the present case, the origin of this effect may be traced back to the magnetic field and its coupling to the hadronic degrees of freedom. Indeed, the place of vanishing Baryonic charge density corresponds to $\cos\theta=0$, making manifest the similarity with the mechanism of charge separation in polarized media. In such media, the electron cloud is distorted by the external field (which moves the center of negative charge relative to the positive ones). Moreover, in both the present case and in polarized media, when the ``external field'' is removed, the charges in the medium generally return to their initial uniform distribution. Here, from the standpoint of the hadronic degrees of freedom, the role of the ``external field'' is played by the Bertotti--Robinson magnetic field, and hence when such a field vanishes, the Baryonic charge density vanishes as well.
The key difference with respect to the previous configuration—where a net Baryonic charge is present—is that, in the Melvin black hole case, the hadronic field is an odd function of the polar angle, whereas in the Bertotti–Robinson case, the hadronic profile is even (while in both cases, the gauge potential has the same parity). We will return to this mechanism of Baryonic charge separation in a future publication.

\section{Conclusions and Perspectives}\label{sec4}

In the present manuscript, we have constructed the first analytic example of a black hole carrying Baryonic charge while immersed in an external delocalized magnetic field. These magnetized Baryonic black holes are either asymptotically Melvin--Bonnor or asymptotically Bertotti--Robinson. Nicely, we have found that the mass and the Baryonic charge are not independent parameters: we derived a closed analytic expression relating the black hole mass parameter to the (discrete) Baryonic charge and the magnetic field. As this relation is a transcendental equation, we have explicitly analyzed its behavior in relevant limiting regimes, particularly for small masses and moderately large masses.

Our analysis of the Baryonic charge density in the Melvin--Bonnor case shows that the highest density occurs at a finite distance from the horizon and is approximately distributed over a spherical-shell-like region. This is a remarkable outcome for several reasons. First, the Baryonic charge is a discrete topological charge rather than a Noether charge, and to our knowledge, there are no explicit examples in the literature where a black hole mass parameter is expressed directly in terms of a discrete charge. In the present case, the connection between the Baryonic charge and the mass parameter follows from the quantization condition imposed on the Baryonic charge itself. Second, even in flat spacetime and neglecting gravitational backreaction, it is uncommon in strongly interacting systems—particularly at large Baryonic charge and in the presence of intense magnetic fields—to obtain an analytic determination of how the total energy depends on the Baryonic charge and the magnetic field.

On the other hand, in the Bertotti--Robinson case, the Baryonic charge density distribution mimics the electric charge distribution of a neutral object polarized by an external agent; indeed, the total net Baryonic charge vanishes. This effect clearly warrants further investigation.

This framework opens several new directions. In particular, it allows the study of the magnetic susceptibilities of the present Baryonic Melvin black hole through the derivatives of the mass parameter with respect to the magnetic field. This is especially appealing since the magnetic susceptibility encodes key information about the underlying phase structure. This, in turn, opens up the possibility for a detailed exploration of the thermodynamic properties of these backreacting Baryonic configurations.

It would be both interesting and instructive to investigate other exact solutions of the Einstein--Scalar--Maxwell theory that generate nontrivial Baryonic charge configurations, thereby enabling a more systematic study of the phase structure and thermodynamics of these novel systems.

\acknowledgments
\vspace{-0.6cm}
The work of J.B. is supported by FONDECYT Postdoctorado grant 3230596. F.C. has been funded by FONDECYT grants 1240048, 1240043, 1240247, and by grant ANID EXPLORACIÓN 13250014. 
A.C. is partially supported by FONDECYT grant 1250318. The work of K.M. is funded by Beca Nacional de Doctorado ANID grant 21231943.
A.N. is supported by ANID-Subdirección de Capital Humano/Doctorado Nacional/2025-21253071.
The Centro de Estudios Cientificos (CECs) is funded by the Chilean Government through the Centers of Excellence Base Financing Program of ANID. The authors would also like to thank the Conference ``Black Hole Physics: Radiation, accelerated spacetimes and beyond'' in Lic\'an Ray for providing such an amazing environment in which this work was completed.

\vspace{-0.4cm}
\bibliography{biblioMELVIN}

\end{document}